# Terahertz dual phase gradient metasurface: high-efficiency binary-channel spoof surface plasmon excitation

LI-ZHENG YIN, TIE-JUN HUANG, DI WANG, JIANG-YU LIU AND PU-KUN LIU *

*State Key Laboratory of Advanced Optical Communication Systems and Networks, Department of Electronics, Peking University, Beijing 100871, China*
*Corresponding author: pkliu@pku.edu.cn

**Spoof surface plasmon meta-couplers are compact antennas which link propagating waves and surface waves. However, most of them are designed with a fixed phase gradient and channel for the incident waves with specific polarization, which limits their further applications in multichannel scenarios. In this Letter, we propose, to the best of our knowledge, a new method that combines the Brillouin folds theory with the Generalized Snell Law. We demonstrate that when the phase gradient of the metasurface is large enough, Brillouin folds effect will occur, which will create dual phase gradient space in a single metasurface. With this method, we design two novel terahertz meta-couplers with functionalities of symmetrical and asymmetrical binary-channel/bidirectional SSP excitation. Furthermore, finite element method (FEM) simulations are performed to demonstrate their functionalities. Considering the orthogonality of the incident waves, there can be a total of four independent space channels to excite SSP on one metasurface. This work may open up new routes in multi-channel SSP meta-couplers and multi-beam surface wave antennas.**

Spoof surface plasmon (SSP), which is the replacement of surface plasmon polariton (SPP) in the microwave and terahertz regimes, possesses intriguing properties of subwavelength confinement and local field enhancement. Thus, it becomes ubiquitous in diverse areas ranging from perfect absorbers [1], and sub-diffraction imaging [2-4] to microwave waveguides [5, 6], etc. The existence of SSP has been proved in structured periodic arrays such as metal gratings. However, due to the large wave vector mismatch between the SSP and free space waves, SSP cannot be excited directly by simply impinging electromagnetic waves (EMWs) on the SSP waveguides. To compensate for the wave vector mismatch, many SSP couplers have been proposed to introduce additional wave vectors, such as depth gradient metal gratings[7, 8].

Recently, meta-couplers, metasurfaces made up of periodic elements with gradient reflection phase that are capable of introducing arbitrary additional wave vectors, have attracted the progressively increasing attention in the past few years [9-12]. In addition to the miniaturization (the thickness is far smaller than the wavelength), compared with the conventional SSP couplers, meta-couplers also have advantages in their potentials in designing multichannel and multifunctional SSP coupling devices[9, 13-19]. In 2013, circular polarization-dependent bidirectional surface plasmon polaritons excitation based on geometric-phase metasurfaces was firstly proposed in the optical regimes [19]. The high-efficiency counterpart was later demonstrated in the microwave regimes by Duan *et al.* [13]. For linear polarization incident waves, the similar functionalities were also numerically and experimentally demonstrated by Meng *et al.* [14]. These works achieve the functionalities of binary-channel/bidirectional SSP excitation by switching the polarization state between the orthogonal incident waves. However, for the incident wave with specific polarization, there is only one channel to excite SSP because the conventional meta-coupler can only provide a fixed phase gradient for all the incident waves, which limits their further applications in multichannel scenarios.

In this work, we propose and demonstrate, to the best of our knowledge, a new method which can realize binary-channel SSP excitation for each specific polarization EMW. According to the theoretical derivation, we prove that dual phase gradient can be formed on a single metasurface by elaborately adjusting the reflection phase of each unit cell and the number of the unit cells in each phase period. Therefore, the incident waves with different incidence angles will "see" different phase gradients and be converted into SSP towards opposite directions. As proofs of the method, two novel terahertz meta-couplers for TM polarization with functionalities of symmetrical and asymmetrical SSP excitation are designed. FEM simulations are performed to demonstrate their functionalities and effects. Furthermore, considering the orthogonality of the incident waves, there can be a total of four channels to excite SSP on a single metasurface. Our work can provide a new way of designing multi-channel SSP meta-couplers and multi-beam surface wave antennas.

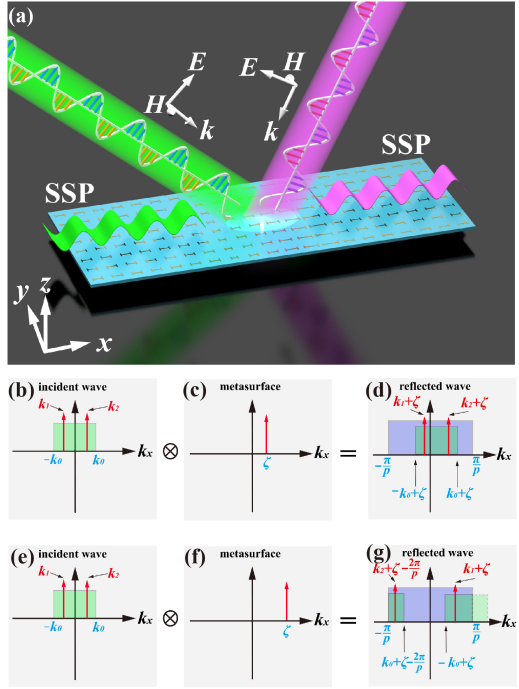

Figure.1. (a) Functional schematic of the dual phase gradient metasurface. (b) Schematic diagram of the Generalized Snell Law. (c) Brillouin folds in the process of the Generalized reflection.

The Generalized Snell Law tells us that the longitudinal wave vector $k_s$ of the reflected/transmitted waves and the incident waves $k_i$ meet the relation

$$k_s = k_i + \frac{\Delta \varphi}{\Delta x}, \tag{1}$$

where $\Delta\varphi$ and $\Delta x$ repersent the phase difference and the distance between the adjacent unit cells of the metasurfaces, respectively [20]. Therefore, $\Delta\varphi / \Delta x$ which is defined as $\zeta$ in this work represents the additional wave vector introduced by the metasurfaces. When $k_s$ is larger than $k_0$, surface wave will be excited on the coupling metasurface, where $k_0$ represents the wave vector in free space [11]. By designing a SSP meta-waveguide whose Eigen wave vector equals to $k_s$, according to the mode coupling theory [21], the surface wave can be efficiently guided out. From Eq. (1) we know that arbitrary $k_s$ can be realized by elaborately adjusting the phase gradient of the metasurfaces and the incidence angle of the incident waves, as can be seen from Figs. 1(b), 1(c), and 1(d). Figure 1(d) is calculated by the convolution of Fig. 1(b) and 1(c). The green shadows in Fig. 1(b) and 1(d) represents the practical wave vector regions of the incident and reflected waves, respectively. In this way, for the incident wave with specified polarization, only the SSP meta-waveguide on one side of the coupling metasurface can be used to guide out the SSP [11]. This is because the additional wave vector introduced by the coupling metasurface is a constant for all the incident waves. As illustrated by Fig. 1(b), and 1(d), for the incident waves with longitudinal wave vectors $k_{i1}$ and $k_{i2}$, the corresponding longitudinal wave vectors of the reflected waves $k_{r1}$ and $k_{r2}$ are

$$k_{r1} = k_{i1} + \zeta, \tag{2a}$$

$$k_{r2} = k_{i2} + \zeta. \tag{2b}$$

To excite SSP in the opposite directions, $k_{r1}$ and $k_{r2}$ should meet the condition

$$|k_{r1} - k_{r2}| > 2k_0 \tag{3}$$

because the Eigen wave vector of SSP is always larger than $k_0$. When substituting Eq. (2) into this inequality, we can get that $|k_{i1} - k_{i2}| > 2k_0$, which cannot be implemented in practice. Therefore, for the EMW with specified polarization, only a single space channel can be used to excite SSP in a predefined direction for the conventional meta-coupler. However, considering the

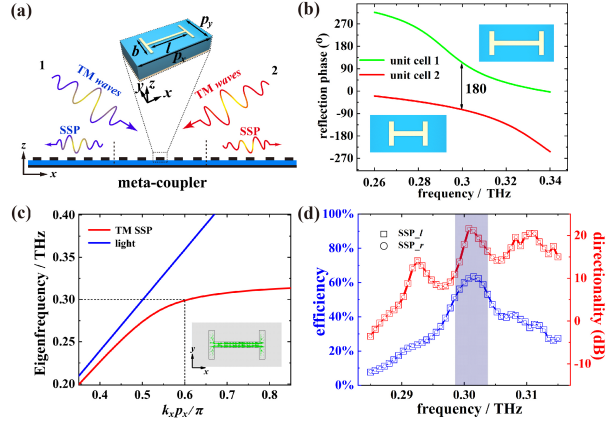

Figure. 2. (a) Schematic of the symmetrical SSP excitation. (b) The reflection phase of the unit cells of the coupling metasurface. (c) The dispersion relation of light and the SSP meta-waveguide. (d) The efficiency and directionality versus frequency of the meta-coupler. The circular and rectangle point plots represent the SSP towards +x and –x directions, respectively.

Brillouin folds effect, the maximum phase difference between the adjacent unit cells is $\pm\pi$, which means that the condition $|k_s| \leq k_{max} = \pi/p$ must be fulfilled, where $p$ represents the period of the unit cells. Therefore, in practice, there is a maximum realizable wave vector region (MRWVR) for each meta-coupler, which are represented by the purple shadows in Figs. 1(d) and 1(g). When $|k_i + \Delta\varphi/\Delta x|$ is larger than $k_{max}$, $2mk_{max}$ will be added on $k_s$ to guarantee that $|k_s| \leq k_{max}$, where $m$ represents an integer. As can be seen from Figs. 1(e), 1(f), and 1(g), when $\zeta$ is large enough, for the region of the reflected wave vector, only a part of it meets the Generalized Snell Law and the rest beyond the MRWVR will be moved to the left. In this case, different additional wave vectors are added on the incident waves with different $k_i$. This phenemenon gives us an opportunity to realize the binary-channel SSP excitation. For the incident waves with longitudinal wave vector $k_{i1}$ and $k_{i2}$, when $k_{i1} + \zeta < k_{max}$ and $k_{i2} + \zeta > k_{max}$ (for arbitrary $k_{i1}$ and $k_{i2}$, there will always be a $\zeta$ that make the two inequalities satisfied), the corresponding longitudinal wave vectors of the reflected waves $k_{r1}$ and $k_{r2}$ meet the relations

$$k_{r1} = k_{i1} + \zeta, \tag{4a}$$

$$k_{r2} = k_{i2} + (1-n)\zeta, \tag{4b}$$

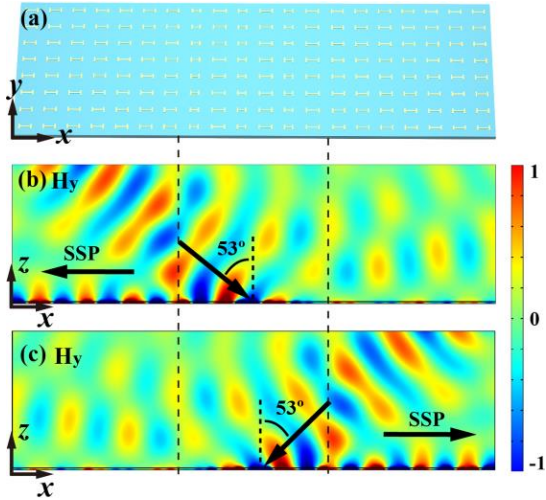

Figure. 3. (a) Schematic of the meta-coupler with the functions of symmetrical SSP excitation. (b) and (c) are respective magnetic field distributions $\mathbf{H}_y$ of the meta-coupler under the illumination of TM polarization EMWs with incidence angles $\theta_i = \pm 53°$.

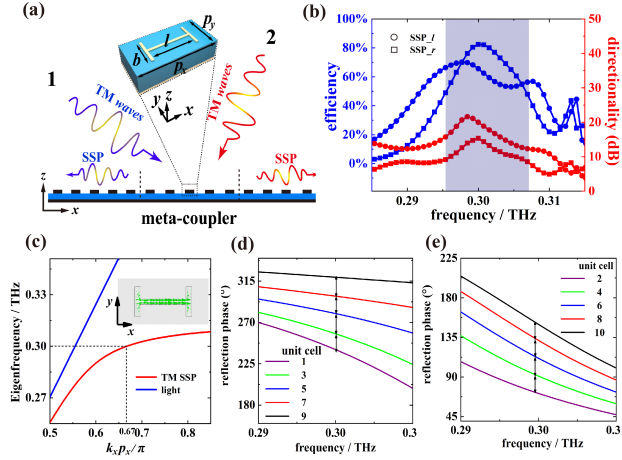

Figure. 4. (a) Schematic of the symmetrical SSP excitation. (b) The efficiency and directionality versus frequency of the meta-coupler. (c) The dispersion relation of light and the SSP meta-waveguide. (d), (e) The reflection phase of the unit cells of the coupling metasurface.

---

where $n = 2\pi / \zeta p$ represents the number of the unit cells in one phase period and it is always larger than two. So the additional wave vectors added on the incident waves have opposite signs. Substituting Eq. (4) into Eq. (3), we get that

$$| k_{i1} - k_{i2} + n\zeta | > 2k_0. \qquad (5)$$

By choosing appropriate $n$ and $\zeta$, Eq. (5) can always be theoretically fulfilled, which means that SSP can be excited in the opposite directions. To demonstrate this method, next, we design two binary-channel meta-couplers with the functionalities of symmetrical and asymmetrical SSP excitation.

First, we design a terahertz meta-coupler with the functionalities of symmetrical SSP excitation, i.e. $k_{i1} = -k_{i2}$ and $k_{r1} = -k_{r2}$, as illustrated by Fig. 2(a). The central working frequency of the meta-coupler is 0.3 THz. In this work, we choose $k_{i1} = -k_{i2} = 0.8\ k_0$ and $k_{r1} = -k_{r2} = -1.2k_0$ so the incidence angles of the two designed space channels are $\theta_i = \pm \arcsin(k_{i1}/k_0) = \pm 53°$. Substituting these design conditions into Eqs. (4a) and (4b), we can get that $n = 2$ and $\zeta = -2k_0$. In this case, the additional wave vectors added on the TM incident waves with incidence angles $\theta_i = \pm 53°$ are $\mp 2\ k_0$. After the theoretical analysis, we design a practical meta-coupler and demonstrate its functions numerically. Figure 3(a) shows the simplified schematic of the meta-coupler. The unit cells of the coupling metasurface and the SSP meta-waveguide have the same shape, as illustrated by the inset in Fig. 2(a). Their length and width are $p_x = 250$ μm and $p_y = 125$ μm, respectively. They are composed of "H" structure copper patches and flat copper mirrors separated by a 40 μm-thick dielectric spacer Taconic RF-43 with the relative permittivity $\varepsilon_r = 4.3 - 0.01i$. The thickness and width of the copper patches and are $h = 2$ μm and t = 16 μm. For the coupling metasurface, the phase difference between the adjacent unit cells is $\Delta\varphi = 360°/n = 180°$ at 0.3 THz. The desired reflection phase can be obtained by changing the shape and size of the metal patches. In the practical simulation process, by adding an ideal periodic boundary condition around the coupling unit cell, the reflection phase can be obtained by calculating the phase of the reflection coefficient of the unit cell. In this way, the optimal structure parameters of the two unit cells in use are $b = 80$ μm, $l = 131$ μm, and $b = 80$ μm, $l = 107$ μm, and their reflection phase versus frequency are plotted in Fig. 2(b). For the unit cells of the SSP waveguide, in the simulation, we apply the ideal periodic boundary condition in the $y$-direction and Floquet boundary condition with longitudinal wave vector $k_f = 1.2k_0$ in the $x$-direction. In this case, we adjust the size of the metal patch so that the Eigen frequency of the SSP waveguide calculated by FEM simulation is equal to the central working frequency of the meta-coupler. In this way, the optimal parameters of the copper pathces are $b = 80$ μm, and $l = 123$ μm and the exact dispersion relation of the SSP waveguide and light are plotted in Fig. 2(c), where the inset illustrates the corresponding electric current distribution of the copper patches at 0.3 THz. Finally, we demonstrate the functionalities of symmetrical binary-channel SSP excitation of the designed meta-coupler. The length of the coupling metasurface is $11 \times p_x = 2.75$ mm. Perfect match layers are set around the meta-coupler to simulate the infinite space. TM-polarized Gaussian waves with waist-width $w$ are selected as the incident wave source. Under these circumstances, the magnetic field distributions $\mathbf{H}_y$ for TM polarization incidence with $w = 0.72\lambda$ and $\theta_i = \pm 53°$ are illustrated by Figs. 3(b) and 3(c), respectively. The visual magnetic field distributions are in agreement with theory, which efficiently confirms the functionalities of binary-channel SSP excitation. To further quantificationally demonstrate the performance of the meta-coupler, the converting efficiency, which is defined as the ratio of the power carried by SSP to that incident on the coupling metasurface, is calculated. The power carried by SSP is calculated by the integral of the Poynting vector. The converting efficiency versus frequency under the condition $w = 1.2\lambda$ is plotted in Fig. 2(d), from which we can find that the respective converting efficiency for the incident waves with $\theta_i = \pm 53°$ can both reach up to 59.8% at 0.3THz. It's noted that the converting efficiencies of the two channels are completely the same because the structure of the meta-coupler is mirror-symmetrical. The gray shadow in Fig. 3(e) represents the frequency band where the converting efficiencies are over 50%. The directionalities, which is defined as the ratio of the power carried by SSP to that flowing towards the opposite directions, are also calculated and plotted in Fig. 2(d). The maximum directionality can reach up to 22 dB at 0.3 THz for both the space channels.

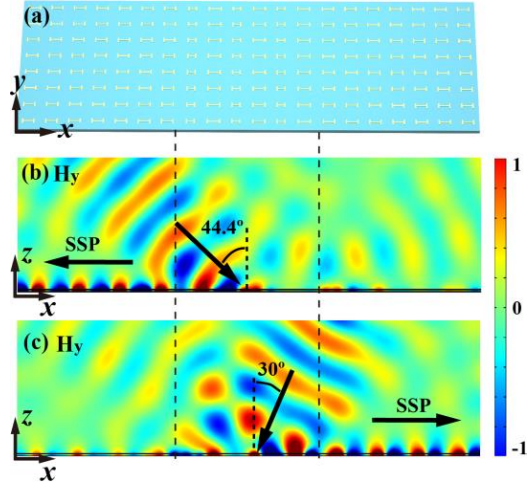

Figure. 5. (a) Schematic of the meta-coupler with the functions of asymmetrical SSP excitation. (b) and (c) are respective magnetic field distributions $H_y$ of the meta-coupler for TM polarization incidence with incidence angles $\theta_i$ = +44.4 ° and -30 °.

To further illustrate the superiority of the proposed method, we design a terahertz meta-coupler with the functionalities of asymmetrical SSP excitation, as illustrated by Fig. 4(a). In this work, we select $k_{i1} = 0.7k_0$, $k_{i2} = -0.5k_0$, $k_{r1} = -1.2k_0$, and $k_{r2} = 1.2k_0$ so the incidence angles of channels 1 and 2 are 44.4 ° and −30 °, respectively. According to Eqs. 4(a), and 4(b), two design parameters can be determined with $n = 2.118$ and $\zeta = 1.7k_0$. The non-integer $n$ means that the coupling metasurface is aperiodic. The unit cells of the coupling metasurface have the same structure and materials with those of the symmetrical meta-coupler and their length and width are $p_x = 277$ μm and $p_y = 139$ μm, respectively. The thickness of the dielectric is 39 μm. The length of the coupling metasurface is $10 \times p_x = 2.77$ mm and the phase difference between the adjacent unit cells is $\Delta\varphi = 360°/n = 170°$ at 0.3 THz. With the same method of obtaining the relation between the reflection phase and $l$ mentioned above, the length of the ten copper patches can be determined with $l = 122.2$ μm, 137.8 μm, 118.3 μm, 135.2 μm, 115.7 μm, 134.7 μm, 110.5 μm, 133.9 μm, 98.8 μm, and 132.6 μm, and their width is $b = 78$ μm. The reflection phase of the designed unit cells versus frequency are also plotted in Figs. 4(d), and (e). To guide out the surface wave, SSP meta-waveguide that can support TM mode SSP with the target Eigen wave vector should be designed. The optimal parameters of the unit cell of the SSP meta-waveguide are $l = 125$ μm, and $b = 80$ μm. The dispersion relation of the meta-waveguide is illustrated in Fig. 4(c) with the electric current distribution of the copper patches inserted below the curves. In the practical simulation process, TM polarization Gaussian waves with $w = 0.86\lambda$ and $1.04\lambda$ impinge on the meta-coupler from channels 1 and 2, and their respective magnetic field distributions are illustrated by Figs. 5(b) and (c). The visual magnetic field distributions can efficiently demonstrate the functionalities of asymmetrical binary-channel SSP excitation. The converting efficiency and directionality, which are the two most important performance indexes, are also calculated and plotted in Fig. 4(b) under the condition $w = 1.43\lambda$ and $1.73\lambda$ for channels 1 and 2. The respective converting efficiency and directionality for channels 1 and 2 are 65.6%, 82.4%, 20dB, and 15 dB at 0.3THz.

In conclusion, a new method which can realize binary-channel SSP excitation for each specific polarization is proposed. As proofs of the method, two terahertz meta-couplers with functionalities of symmetrical and asymmetrical SSP excitation are designed. FEM simulations are also performed to demonstrate their functionalities and effects. This work can provide a new way of designing multi-channel SSP meta-couplers and multi-beam surface wave antennas.

**Funding.** National Key Research and Development Program (2019YFA020015); National Natural Science Foundation of China (NSFC) (61971013).

## References

1. W.-C. Chen, A. Cardin, M. Koirala, X. Liu, T. Tyler, K. G. West, C. M. Bingham, T. Starr, A. F. Starr, and N. M. Jokerst, "Role of surface electromagnetic waves in metamaterial absorbers," Opt. Express **24**, 6783-6792 (2016).
2. T.-J. Huang, J.-Y. Liu, L.-Z. Yin, F.-Y. Han, and P.-K. Liu, "Superfocusing of terahertz wave through spoof surface plasmons," Opt. Express **26**, 22722-22732 (2018).

3. H.-H. Tang and P.-K. Liu, "Terahertz far-field superresolution imaging through spoof surface plasmons illumination," Opt. Lett. **40**, 5822-5825 (2015).
4. T.-J. Huang, L.-Z. Yin, Y. Shuang, J.-Y. Liu, Y. Tan, and P.-K. Liu, "Far-Field Subwavelength Resolution Imaging by Spatial Spectrum Sampling," Phys. Rev. Appl. **12**, 034046 (2019).
5. A. Kianinejad, Z. N. Chen, and C.-W. Qiu, "Low-loss spoof surface plasmon slow-wave transmission lines with compact transition and high isolation," IEEE Trans. Microw. Theory Technol. **64**, 3078-3086 (2016).
6. I. R. Hooper, B. Tremain, J. Dockrey, and A. P. Hibbins, "Massively sub-wavelength guiding of electromagnetic waves," Sci. Rep. **4**, 7495 (2014).
7. T.-J. Huang, L.-Z. Yin, J.-Y. Liu, F.-Y. Han, Y. Tan, and P.-K. Liu, "High-efficiency directional excitation of spoof surface plasmons by periodic scattering cylinders," Opt. Lett. **44**, 3972-3975 (2019).
8. H.-H. Tang, T.-J. Ma, and P.-K. Liu, "Experimental demonstration of ultra-wideband and high-efficiency terahertz spoof surface plasmon polaritons coupler," Appl. Phys. Lett. **108**, 191903 (2016).
9. L.-Z. Yin, T.-J. Huang, F.-Y. Han, J.-Y. Liu, D. Wang, and P.-K. Liu, "High-efficiency terahertz spin-decoupled meta-coupler for spoof surface plasmon excitation and beam steering," Opt. Express **27**, 18928-18939 (2019).
10. J. Wang, S. Qu, H. Ma, Z. Xu, A. Zhang, H. Zhou, H. Chen, and Y. Li, "High-efficiency spoof plasmon polariton coupler mediated by gradient metasurfaces," Appl. Phys. Lett. **101**, 201104 (2012).
11. S. Sun, Q. He, S. Xiao, Q. Xu, X. Li, and L. Zhou, "Gradient-index meta-surfaces as a bridge linking propagating waves and surface waves," Nat. Mater. **11**, 426 (2012).
12. W. Sun, Q. He, S. Sun, and L. Zhou, "High-efficiency surface plasmon meta-couplers: concept and microwave-regime realizations," Light: Sci. Appl. **5**, e16003 (2016).
13. J. Duan, H. Guo, S. Dong, T. Cai, W. Luo, Z. Liang, Q. He, L. Zhou, and S. Sun, "High-efficiency chirality-modulated spoof surface plasmon meta-coupler," Sci. Rep. **7**, 1354 (2017).
14. Y. Meng, H. Ma, M. Feng, J. Wang, Z. Li, and S. Qu, "Independent excitation of spoof surface plasmon polaritons for orthogonal linear polarized incidences," Appl. Phys. A **124**, 707 (2018).
15. S. Liu, T. J. Cui, A. Noor, Z. Tao, H. C. Zhang, G. D. Bai, Y. Yang, and X. Y. Zhou, "Negative reflection and negative surface wave conversion from obliquely incident electromagnetic waves," Light: Sci. Appl. **7**, 18008 (2018).
16. T. Cai, S. Tang, G. Wang, H. Xu, S. Sun, Q. He, and L. Zhou, "High-performance bifunctional metasurfaces in transmission and reflection geometries," Adv. Opt. Mater. **5**, 1600506 (2017).
17. F. Ding, R. Deshpande, and S. I. Bozhevolnyi, "Bifunctional gap-plasmon metasurfaces for visible light: polarization-controlled unidirectional surface plasmon excitation and beam steering at normal incidence," Light: Sci. Appl. **7**, 17178 (2018).
18. Y. Ling, L. Huang, W. Hong, T. Liu, L. Jing, W. Liu, and Z. Wang, "Polarization-switchable and wavelength-controllable multi-functional metasurface for focusing and surface-plasmon-polariton wave excitation," Opt. Express **25**, 29812-29821 (2017).
19. L. Huang, X. Chen, B. Bai, Q. Tan, G. Jin, T. Zentgraf, and S. Zhang, "Helicity dependent directional surface plasmon polariton excitation using a metasurface with interfacial phase discontinuity," Light: Sci. Appl. **2**, e70 (2013).
20. N. Yu, P. Genevet, M. A. Kats, F. Aieta, J.-P. Tetienne, F. Capasso, and Z. Gaburro, "Light propagation with phase discontinuities: generalized laws of reflection and refraction," Science **334**, 333-337 (2011).
21. K. Zhang, D. Li, K. Chang, K. Zhang, and D. Li, *Electromagnetic theory for microwaves and optoelectronics* (Springer, 1998).